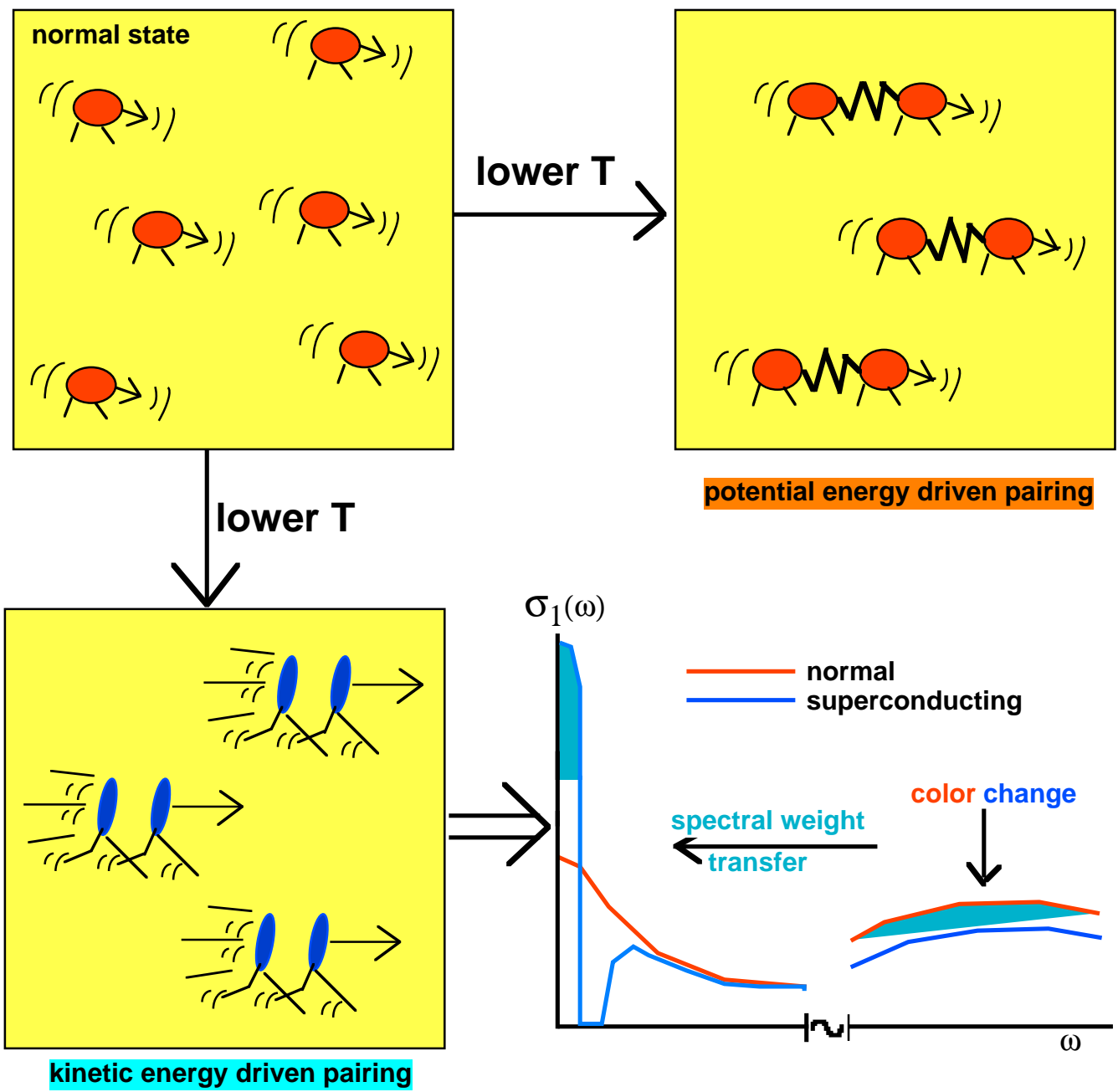

Two routes to pairing and superconductivity. In the conventional route (upper right), carriers lower their potential energy upon pairing. In the new route (lower left), carriers lower their kinetic energy upon pairing, they become lighter and can propagate more easily. In both cases optical spectral weight at low frequencies is rearranged due to the opening of the superconducting energy gap. Only in the latter case is optical spectral weight transfered from the high frequency visible region to the low frequency region.

# The True Colors of Cuprates

J.E. Hirsch

In the early days of superconductivity researchers expected that superconductors, by analogy with good metals, would be particularly good reflectors of high frequency radiation. However, experimental attempts (1) to detect changes in reflectivity in the visible range when metals were cooled below their superconducting transition temperature ($T_c$) found no effect whatsoever. Half a century later, Molegraaf et al report in p. 2239 of this issue that at least some superconductors indeed appear to follow this expectation.

This finding is especially significant because the search for such an effect was completely abandoned after 1957, when the BCS (Bardeen-Cooper-Schrieffer) theory of superconductivity was developed (2). BCS theory has two key ingredients: one is the realization that superconductivity results when electrons pair up into 'Cooper pairs', thus turning from fermions to bosons that can condense into a coherent state that is insensitive to impurities and imperfections in the metal and hence conducts electricity without resistance; the second is that the pairing is mediated by the interaction of electrons with 'phonons', the quanta of ionic vibrations of the crystal.

Because electrons form pairs with binding energy $2\Delta \sim k_B T_c$, photons of energy $h\omega < 2\Delta$ cannot be absorbed and optical absorption is inhibited at low frequencies (far infrared and microwave); indeed, at the same time that BCS theory was being developed the suppression of low frequency optical absorption in superconductors was being detected experimentally by Glover and Tinkham (3). The interaction between electrons and phonons, which mediates the Cooper pair formation in BCS theory, can give rise to changes in optical absorption at frequencies $2\Delta + h\omega_0$, with $\omega_0$ a phonon frequency (Holstein processes), in the infrared range. However, no mechanism exists in BCS theory for changes in optical absorption at much higher frequencies, in the visible range. Since BCS theory was rapidly accepted as the correct explanation of superconductivity, due to its logical consistency and its ability to explain a vast array of experimental observations, no further attempts to measure superconductivity induced 'color changes' were undertaken for many years.

Enter the high $T_c$ cuprates. These remarkable materials discovered in 1986 (4) exhibit superconductivity at temperatures much higher than their conventional counterparts. The fact that charge carriers in high $T_c$ cuprates form Cooper pairs when they superconduct is not in doubt: it is supported by a variety of experimental observations, and no other mechanism to obtain superconductivity is known. However, the mechanism by which carriers pair, that is responsible for the high transition temperatures, has been the subject of intense study and heated debate for the past 15 years, and no consensus has been reached. It is generally agreed that the conventional electron-phonon pairing mechanism cannot explain the superconductivity of the cuprates because the energy scale of lattice vibration frequencies is far too small to explain transition temperatures as high as 160K as seen in the cuprates.

In the conventional BCS pairing theory the potential energy of a pair is lowered through interaction with the phonons. Even though a small cost (i.e. increase) of kinetic energy occurs, the larger lowering of potential energy wins out and provides the condensation energy that stabilizes the superconducting state. In the theoretical search for the explanation of pairing in the cuprates, new mechanisms were envisaged where in fact the carriers would lower their <u>kinetic</u> energy, even if at a cost in potential energy, when they form a pair (5,6). It was also realized that kinetic-energy-driven superconductors would exhibit qualitatively new features in their optical properties: (i) the 'violation' of a low frequency optical sum rule (7,8), and (ii) a 'change in color' (9), i.e. change in high frequency optical absorption, when the material becomes superconducting.

The kinetic energy of carriers depends on their effective mass $m^*$. When carriers are heavily 'dressed' by interactions with other carriers or other degrees of freedom their effective mass is large and they cannot respond well to low-frequency electric fields; hence the low frequency (intraband) optical absorption is small, and most of the optical absorption occurs at high frequencies. If $m^*$ decreases when carriers pair, they are better able to respond to low frequency electric fields and optical spectral weight is transferred from high frequencies in the visible range to low frequencies in the intraband range; the decrease in high frequency absorption is the 'color change'. As $m^*$ decreases the carriers can delocalize better and the kinetic energy decreases: in quantum mechanics, localization is associated with higher kinetic energy due to quantum-mechanical zero point motion.

Optical experiments by Basov and coworkers (10) in 1999 detected a small amount of kinetic energy lowering in the c-axis (interlayer) transport of high $T_c$ cuprates by examination of the low frequency optical sum rule. This finding provided some support for the interlayer tunneling mechanism (ILT) of Anderson and coworkers (5), whereby the condensation energy would arise from interlayer pair delocalization. Unfortunately, careful experiments by Moler et al (11) and by Tsvetkov et al (12) had already shown that interlayer tunneling could provide no more than 1% of the condensation energy in certain cuprates and, as a result, ILT is no longer considered a viable mechanism for superconductivity in the cuprates (13).

The experimental results reported by Molegraaf et al show convincingly that changes in the optical conductivity occur in the visible range of the spectrum, that track the opening of the superconducting energy gap. Molegraaf et al do not directly measure the expected violation of the low energy sum rule that should accompany this effect; however, very recent measurements by Santander et al (14) in the infrared range show precisely that effect. Both Molegraaf et al and Santander et al are able to quantify the amount of kinetic energy lowering implied by their experimental results at about 1meV per Cu atom. This is much larger than the amount required to account for the condensation energy of this material (about 100 μeV per Cu atom) (15); this is however not surprising, since the short coherence length in the cuprates suggests that the cost in Coulomb repulsion between members of a Cooper pair should be substantial.

These experimental results place very strong constraints on theories of high temperature superconductivity, and hence will play an important role in the elucidation of its cause. A successful theory needs to explain the physical origin of the kinetic energy lowering and be consistent with the magnitude, temperature and doping dependence of the kinetic energy lowering implied by the data; it needs to explain the physical origin of the high energy scale that gives rise to the observed optical changes in the visible range, and elucidate the mechanism by which the remarkable coupling between this high energy scale and the low energy scale of the superconducting energy gap occurs. In particular, theories that adscribe the pairing mechanism to a magnetic exchange coupling J $S_1 S_2$ predict lowering of exchange energy and increase of kinetic energy upon pairing (16) and hence would appear to be qualitatively inconsistent with these findings. Theories based on stripes (17) and on spin-charge separation (18) may be consistent with these findings and need to be reexamined in light of these results; even though it has been proposed that kinetic energy lowering will result in these theories, its expected magnitude needs to be quantified and the physical origin of the high energy scale needs to be clarified. In the theory of hole superconductivity (6) the high energy scale arises from coupling of the hole to electronic excitations; pairing gives rise to 'undressing' of hole carriers, and various predicted consequences (19) appear to be consistent with the reported observations. That theory, unlike others, predicts that these phenomena should occur in other superconductors besides high $T_c$ cuprates (20). New theories based on kinetic energy lowering will undoubtedly be formulated stimulated by these findings.

A new and qualitatively different paradigm for superconductivity is emerging. Carriers pair not because they are happy 'being' together, as in the BCS paradigm, but because they are happy 'moving' together, even if uncomfortable in each other's presence. It is like car-pooling with somebody you don't like. The results of Molegraaf et al show that kinetic-energy-driven superconductors, heretofore a theoretical construct, indeed exist. Because what defines a superconductor is precisely the enhanced ability of the carriers to propagate and thereby conduct electricity, this new paradigm makes eminent sense.